# Hot-electron bolometric mixer with negative differential resistance

Chang Yoo [1,3], Akim A. Babenko [1,2], and Boris S. Karasik [1,a)]

**Abstract — We demonstrate that the conversion gain of a superconducting hot-electron bolometer (HEB) mixer can be increased by biasing the device within the negative differential resistance (NDR) region of its current–voltage characteristic. Although NDR biasing has historically been avoided due to MHz-range resistive oscillations, we show that these oscillations arise from an LC resonance formed by the bias-T inductance and the effective thermal capacitance of the HEB. By applying stability criteria analogous to those developed for tunnel diodes, we redesigned the embedding circuit to suppress this resonance and achieve stable NDR operation. Direct measurements using two monochromatic 2.5-THz sources confirm the predicted gain enhancement. These results establish NDR biasing as a viable method for improving HEB mixer performance and motivate further studies of noise behavior and circuit optimization.**

**Affiliations**:

[1] Jet Propulsion Laboratory, California Institute of Technology, Pasadena, CA 91109, USA

[2] Currently, with Google Quantum AI, Goleta, CA 93111, USA

[3] Currently, with xLight, Palo Alto, CA 94306, USA

[a)] **Author to whom correspondence should be addressed:**

boris.s.karasik@jpl.nasa.gov

Superconducting resistive bolometers are essential for remote sensing, spectroscopy, and quantum information applications. Although they require cryogenic cooling, they outperform non-superconducting sensors in many high-sensitivity regimes. Depending on the readout method, these devices operate either as direct detectors—transition-edge sensors (TES)—or as heterodyne mixers, known as hot-electron bolometers (HEBs). TES devices, typically fabricated from low-$T_C$ materials such as Ti, W, Al, Mo, and their bilayers with Au or Cu, use these materials as thermometers. Their heat capacity and thermal conductance to the heat sink are controlled by mechanical design, often employing $Si_xN_y$ membranes. This approach allows independent tuning of TES parameters, enabling low thermal conductance $G$ for high responsivity and reduced heat capacity $C$ for millisecond-scale thermal time constants $\tau_{th} = C/G$ [1].

Another TES approach uses the hot-electron effect, where thermal conductance is dominated by weak electron-phonon coupling [2]. In thin-films configuration, thermal phonons escape without returning energy to the electrons, so the effective thermal conductance is set by electron-phonon interactions, and the thermal time constant corresponds to the intrinsic electron-phonon scattering time $\tau_{th} = \tau_{e-ph}$.

HEB mixers [3,4], which are critical for astrophysics in the 1.3-6 THz range, operate on similar principles, but in practical HEB devices the phonon escape time is non-negligible, leading in combination with $\tau_{e-ph}$ to thermal time constants of several tens of picosecond. This is necessary to achieve intermediate frequency (IF) bandwidth of several GHz. As a result, only NbN films with a thickness of 3–5 nm and $MgB_2$ films of 5–10 nm have demonstrated successful HEB mixer performance.

TES performance improves under voltage bias due to negative electro-thermal feedback (n-ETF), which stabilizes the bias point. Optimal operation occurs in the negative differential resistance (NDR) region, where the device exhibits a faster response, higher responsivity, and reduced Johnson and readout noise [5]. Most TES devices remain stable because the electrical time constant $\tau_{el} = L_{SQUID}/(R + R_B) \sim$ 0.1–1 μs is much smaller than the thermal time constant $\tau_{th} \sim$ ms ($L_{SQUID}$ is the SQUID readout input coil, $R_B \leq 1$ Ohm is the bias resistor, $R$ is the TES ohmic resistance). Analyses [6,7] confirm these conditions satisfy stability requirements.

In principle, n-ETF could similarly improve HEB mixer performance by increasing conversion gain [8] and reducing the system noise temperature. However, this approach has not yet been implemented in HEB mixers because the thermal time constant is very short ($\tau_{th}$~ tens of picoseconds) and the external electrical embedding circuit is comparatively complex.

In superconducting HEBs, the NDR regime can be accessed by lowering the local oscillator (LO) power, which produces an N-shaped current-voltage characteristic (IVC). However, operating in this regime often triggers MHz-range oscillations preventing stable mixer operation [9-14]. These oscillations have been attributed to magnetic vortex motion [10-12], but circuit analysis shows that NDR amplifies resonances caused by reactive components in the bias network. In such cases, the oscillation frequency is determined by the LC resonance rather than by vortex dynamics.

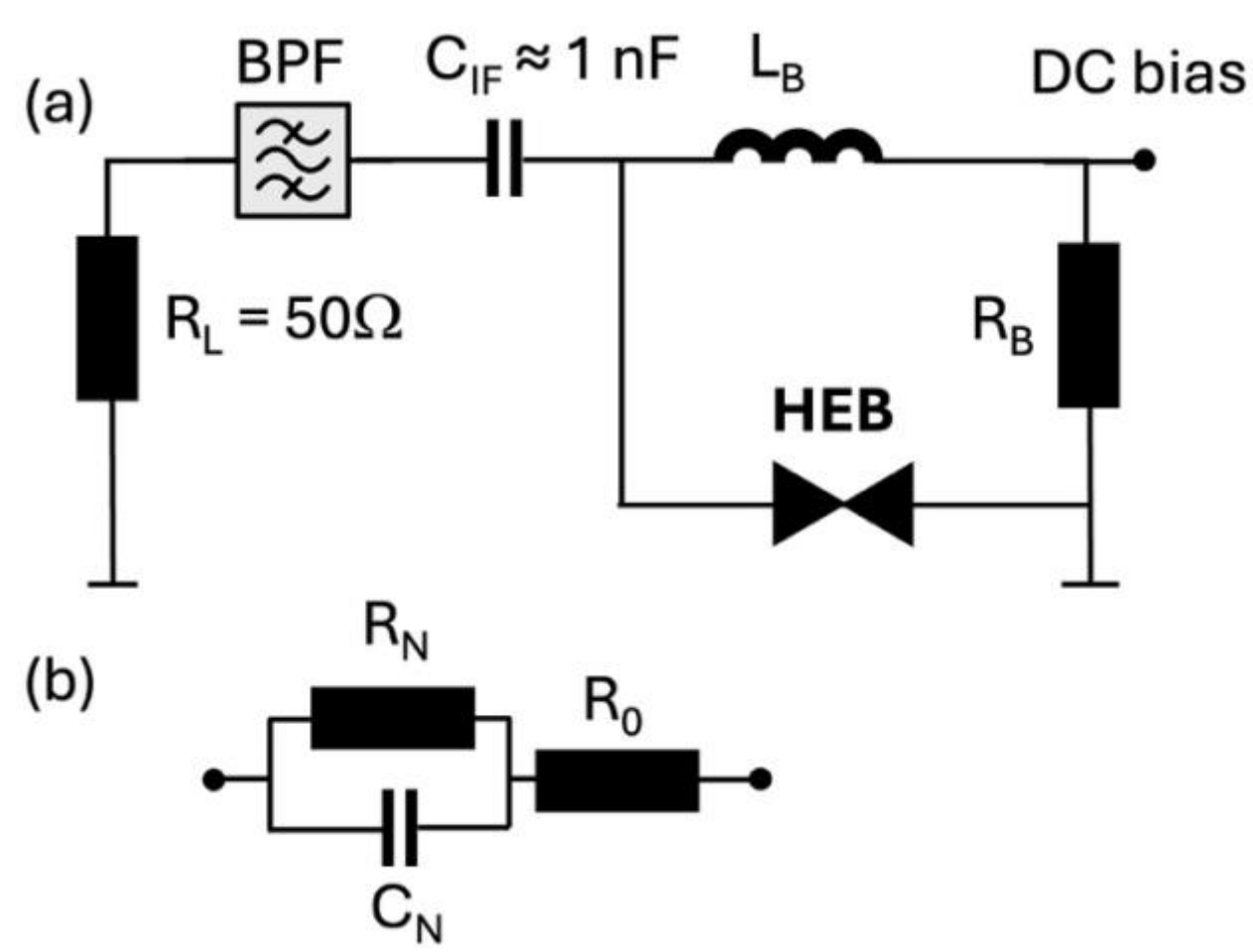

**Figure 1.** (a) Schematic of a typical HEB bias/IF circuit. In our experiment, $R_B$= 15 Ohm, $L_B$= 2 μH or 220 nH, or 70 nH. The resulting equivalent bandpass filter (BPF) bandwidth was 0.8-2 GHz. (b) Equivalent electrical circuit of the HEB mixer. The component values correspond to the parameters of the bolometer (see Table I).

Similar phenomena have been observed in other passive solid-state NDR devices, including tunnel diodes and Gunn diodes. Methods for stabilizing the bias point in resonant tunneling diodes (RTDs) have been studied since the 1960s, and the resulting criteria apply broadly to any NDR device [15-20].

The objective of the current work was to resolve the instability issue in HEB mixer devices with NDR and demonstrate the mixer operation with an increase in conversion gain. We consider the typical HEB bias/IF circuit including the bias-T inductance $L_B$, the IF coupling capacitor $C_{IF}$, and the DC bias resistor $R_B$. The IF line includes an L-band isolator and a low-noise amplifier (LNA) modeled as a 50-Ohm termination (Fig. 1(a)). Because the IF bandwidth (0.8–2 GHz, set by a combined response of the LNA and the L-band isolator) is far above the parasitic oscillation frequencies, it is excluded from the stability analysis.

The HEB device is modeled as two resistors ($R_0$ and $R_N$) in series, with $R_N$ being shunted by a capacitor $C_N$ (Fig. 1(b)) This representation follows the bolometric impedance model for the HEB [8,21,22] and has been experimentally validated [8,23-25]. The resistor $R_N$ represents the inertial part of the differential resistance that vanishes at frequencies above the thermal response speed $\tau_{th}^{-1}$. The resistor $R_0$ is a real part of the bolometer impedance that remains constant over a broad frequency range $\tau_{th}^{-1} < \omega < 2\Delta/\hbar$ where $\Delta$ is the maximum superconducting gap in the resistive state. In most cases, $R_0 = R = V/I$ [8,25], although a moderate current dependence $R(I)$ has been observed in the vortex-driven resistive [23] state. In the following, we assume $R_0 = R$. The capacitance $C_N$ represents the equivalent of the thermal inertia.

The IF impedance of the HEB is given by the lumped-element bolometric theory [21,23]:

$$Z(\omega) = R\frac{1+\mathcal{L}}{1-\mathcal{L}} \cdot \frac{1+j\omega\tau_{th}/(1+\mathcal{L})}{1+j\omega\tau_{th}/(1-\mathcal{L})}. \quad (1)$$

Here $\mathcal{L}$ is the dimensionless parameter called the "self-heating parameter" in the HEB community and "the electro-thermal feedback loop gain" in the TES community. We use the standard TES notation to distinguish it from other notations in this paper. $\mathcal{L} \equiv dP_J/(GdT_e)$ is always positive and depends on the shape of the IVC ($P_J$ is the Joule power dissipated at the bias point). For the NDR case where $Z(0) < 0$, $\mathcal{L} > 1$.

The equivalent circuit of Fig. 1(b) [8] yields:

$$Z(\omega) = R_0 + \frac{R_N}{1+j\omega R_N C_N}. \quad (2)$$

Comparison of Eqs. (1) and (2) yields the relationships summarized in Table I.

**TABLE I. Relationship between the electrical model and bolometric physical parameters.**

| Equivalent circuit parameter | Relation to the bolometric parameters |
|---|---|
| $R_0$ | $R\ (\equiv V/I)$ |
| $R_N$ | $2\mathcal{L}R/(1-\mathcal{L})$ (< 0 for NDR) |
| $Z(0) \equiv dV/dI = R_0 + R_N$ | $R\,(1+\mathcal{L})/(1-\mathcal{L})$ (< 0 for NDR) |
| $C_N$ | $\tau_{th}/(2R\mathcal{L})$ |

We experimented with three HEB devices. Only one NbN device (#1), whose parameters are listed in Table II, underwent the complete investigation cycle, and the following discussion and analysis are based on the data obtained from this device. For two other devices (another NbN device, #2, and an $MgB_2$ HEB device, #3, made from a sputtered film [26]), only partial data showing instabilities in the NDR regime were obtained.

**TABLE II. Characteristics of NbN HEB device #1**

| Parameter | Value |
|---|---|
| Film thickness | 5-6 nm |
| Substrate | Si with MgO buffer |
| Normal resistance | 104 Ohm |
| Critical temperature | 11.1 K |
| Superconducting transition width | 2.1 K |
| Critical current at 4.2 K | 240 μA |

We initially used a commercial bias-T with an inductance $L_B \approx 2$ μH (room temperature value) and a coupling capacitor $C_{IF}$ = 1 nF. With this circuit, we carried out experiments using a common quasioptical HEB mixer design, where the mixer device integrated with a log-spiral antenna on a silicon chip is placed on a hyperhemispherical Si lens for coupling to THz radiation. A 2.52-THz methanol gas laser served as local oscillator (LO), and a weak signal from a frequency multiplier chain (FMC) offset by 1.4 GHz was injected into the mixer through the same port as the LO to monitor the intermediate frequency (IF) output at 1.4 GHz, which was proportional to the mixer conversion gain.

With this circuit, the NDR region was unstable. Figure 2 (blue symbols) shows a plateau with a kink rather than a smooth NDR region. Instead of a single 1.4 GHz peak, the IF spectrogram (inset in Fig. 2) displayed intermodulation sidebands spaced by 10-20 MHz, depending on the bias point. This behavior is consistent with the expected LC resonance frequency $f_{osc} \approx 1/\left(2\pi\sqrt{L_B C_N}\right)$ ~ few 10s MHz ($C_N$ ~ a few ps). Devices #2 and #3 showed almost identical intermodulation spectra. Although device #3 was fabricated from a different superconducting material ($MgB_2$), its thermal time constant (≈ 60 ps [26]), an consequently $C_N$ value, was similar to those in NbN devices (#1 and #2). As a result, their corresponding $f_{osc}$ values were also similar. The similarity of the $f_{osc}$ values in all three devices provides additional evidence that the cause of oscillations lies in the circuit rather than in the magnetic vortex dynamics.

Large-amplitude switching between two IVC points also occurred, producing time-averaged distortions in the measured IVC. This effect was reported in many HEB papers (see, e.g., [27,28]) but has never been properly explained. However, an identical behavior of the unstable IVC with NDR was studied and explained for tunnel diodes [29,30]. Another evidence of the spurious nature of the IVC with NDR is the conversion gain behavior. When the bias is swept along the stable IVC (see Fig. 2) the gain changes smoothly and monotonically. However, in the NDR range the gain drops dramatically. This because there is no real stationary IVC and the observed gain values are just time averaged random samples captured by the spectrum analyzer.

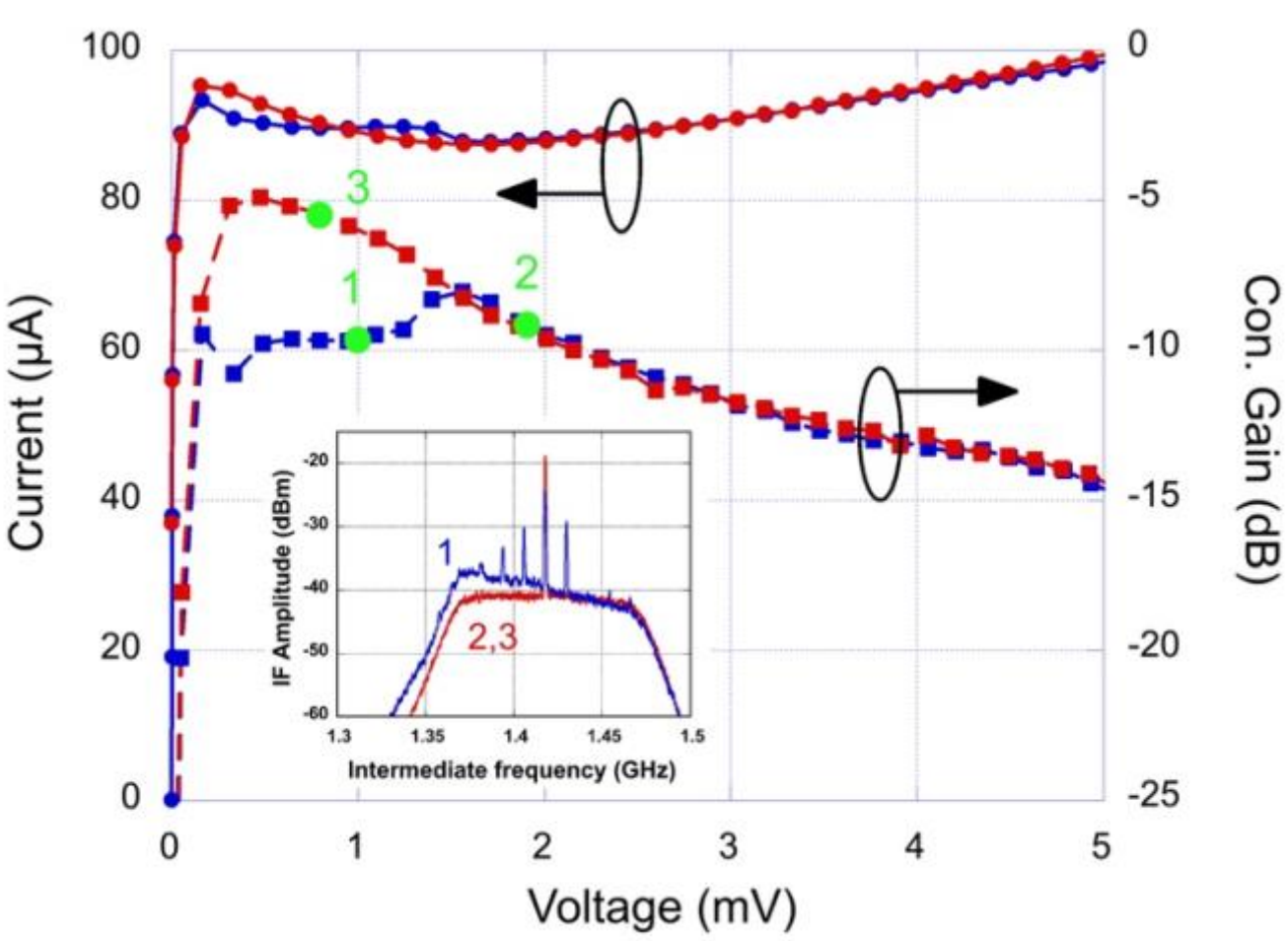

**Figure 2.** IVCs and corresponding conversion-gain curves vs bias for device #1. Blue symbols correspond to $L_B$= 2μH and red symbols are for $L_B$= 220 nH. The plateau on the "blue" IVC below 1.5 mV is the result of time-average oscillations caused by the instability of the bias point. The conversion gain values in this region are also uncertain and deviate from the steady trend above 1.5 mV. The inset shows the intermodulated components (10-20 MHz) caused by the bias point oscillation around the mixing tone at ≈ 1.4 GHz (point 1). In the steady states (points 2 and 3), these components are absent. The 100-MHz output bandwidth shown in the inset was determined by the room-temperature bandpass filter connected to the receiver IF output. It is much narrower that the IF bandwidth presented to the HEB device (0.8-2 GHz).

The HEB circuit resembles the circuit used to analyze the Esaki tunnel diode closely. The stability of tunnel diodes with NDR was studied intensively by solving the circuit differential equations and identifying the solution leading to damped or overdamped oscillations [19]. In the terms used in this paper, the general criteria can be summarized into two conditions [17]:

$$R_{DC} < |Z(0)|, Z(0) = dV/dI \tag{3a}$$

$$\omega_r < \omega_x \tag{3b}$$

Here $R_{DC} = R_0 + R_B$ is the total resistance in series with the pure negative $R_N$ . $\omega_r = (R_N C_N)^{-1}\sqrt{R_N/R_s - 1}$ is the frequency point where the real part of the total circuit impedance turns zero, $\text{Re}[Z_{tot}(\omega_r)] = 0$. Similarly, $\omega_x = \sqrt{(L_b C_N)^{-1} - (R_N C_N)^{-2}}$ yields $\text{Im}[Z_{tot}(\omega_x)]= 0$. Equation (3a) states that the load line's DC resistance, $R_{DC}$, must be less than the absolute value of the NDR, $|Z(0)|$; that is, the load line intersects the IVC at only one point (so-called "voltage bias", in TES terminology). This prevents the device from switching betw een points on the IVC where $Z(0) > 0$ and the bias is stable, across the NDR bias range where $Z(0) < 0$. Equation (3b) states that the characteristic frequency of the resonance should be away from the range where $\mathrm{Re}(Z) < 0$ so the oscillations are damped.

Using Table I, condition 3(b) can be reduced to the following:

$$L_B < \tau_{th}(R_B + R)/(\mathcal{L} - 1). \quad (4)$$

Although the lumped-element model describes NbN HEB only qualitatively due to their hot-spot nature [31,32], Eq. (4) provides useful guidance: $L_B$ must be minimized, and $R_B$ should be as large as possible while still satisfying Eq. (3a).

Following Eq. (4), we replaced the commercial bias-T with a custom-made one, using high-frequency Piconics™ air-core inductors $L_B$ = 220 nH or 70 nH. The bias resistance remained unchanged ($R_B$ = 15 Ohm). This made the IVC continuous (Fig.2, red symbols), and the IF spectrum reduced to a single tone. Further reduction of the LO power (smaller $|Z(0)|$ and larger $\mathcal{L}$ values) led again to instabilities but the frequency of the intermodulation sidebands became of the order of 100 MHz, roughly corresponding to a lower $L_B$ value.

Figure 3 shows the gradual transition of the PDR into NDR when the LO power is decreased in small steps and $L_B$ = 70 nH. The behavior of the conversion gain is similar to that in Fig. 2: for the IVCs with low pumping level, the NDR regions exhibited kinks, and the apparent conversion gain was low and choppy, as the bias was swept. Starting from curve 5, the NDR bias was stable. The highest gain was achieved for curves 5 or 6 (almost no difference), maybe slightly higher than in the case of $L_B$ = 220 nH, since the critical current (and $\mathcal{L}$ value) was higher.

The gain in the NDR regime is a few dB higher than in the PDR regime (Fig. 3). The moderate magnitude of this effect is mostly due to the mixer bandwidth compression for $\mathcal{L}$ values only slightly greater than 1 that was possible to achieve in this experimental work. At least qualitatively, the gain and the mixer time constant can be explained by the lumped-element model equations:

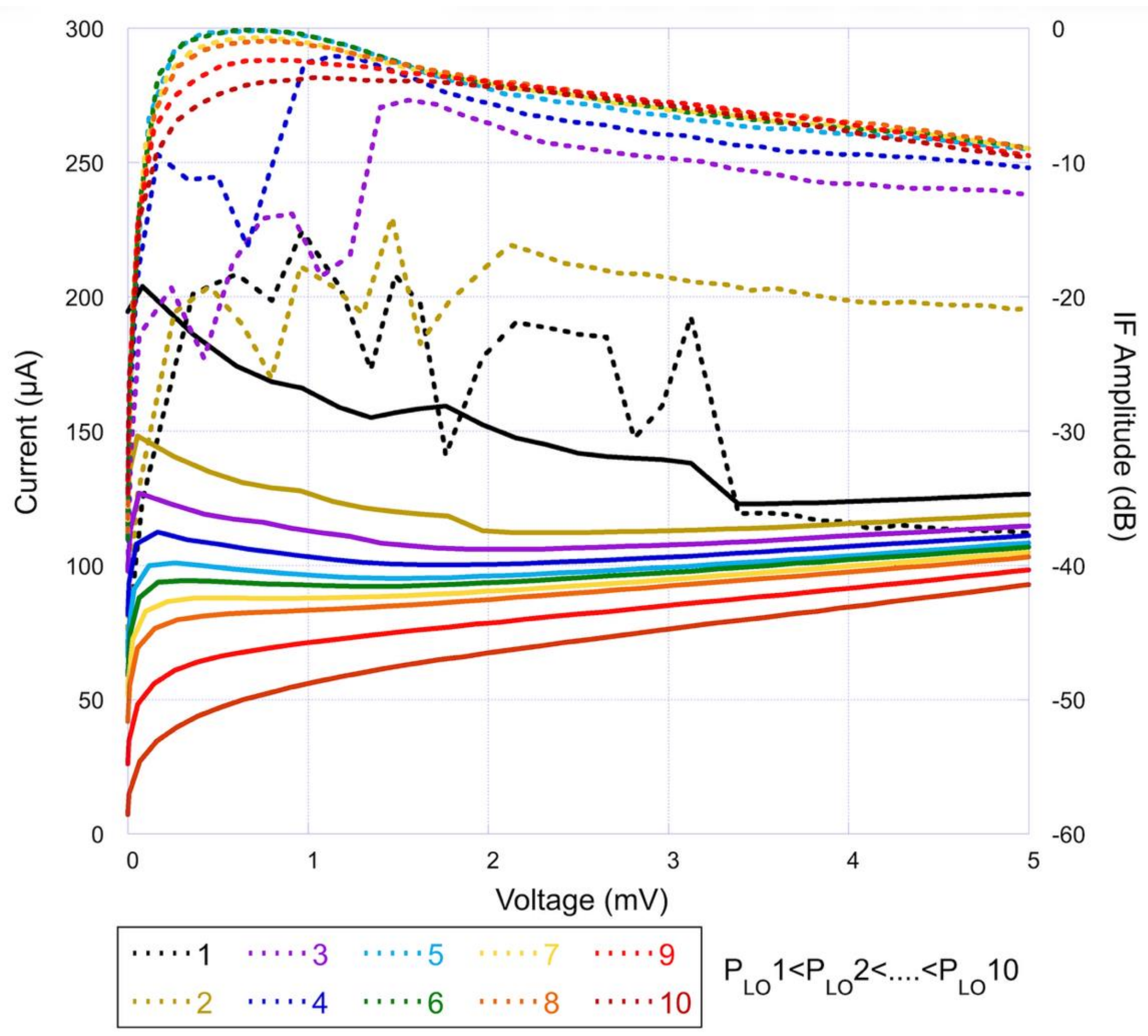

**Figure 3.** Gradual transition from NDR to PDR in device #1 as the LO power decreases (curves numbers from 1 to 10), and $L_B = 70$ nH. Solid lines are IVCs, dashed lines are IF output power (proportional to the conversion gain). Same line colors correspond to the same $P_{LO}$ values. Gain curves 5 and 6 are almost indistinguishable.

$$\eta = 2\mathcal{L}^2\left(P_{LO}/P_J\right)\frac{RR_L}{(R+R_L)^2}\frac{1}{\left(1+\mathcal{L}\frac{R-R_L}{R+R_L}\right)^2}\frac{1}{1+(\omega_{IF}\tau^*)^2} \tag{5a}$$

$$\tau^* = \tau_{th}/\left(1+\mathcal{L}\frac{R-R_L}{R+R_L}\right) \tag{5b}$$

Here $\omega_{IF}$= 2π ×1.4 GHz and $\tau_{th}$≈ 60 ps for our 5-6 nm thick film[33,34], $R_L$=50 Ohm is the IF line impedance. In our case, $R \approx 5$ Ohm (the peak gain value in Figs. 2 and 3), the maximum $\mathcal{L}$ value is ≈ 1.1 as estimated from the experiment IVC and Eq. (1) assuming $\omega_{IF}$= 0. Then $\tau^* \approx 10\times\,\tau_{th}$ and the overall gain $\eta$ is a factor of ≈ 14-15 dB lower than it would have been for $\omega_{IF} \approx 0$. For HEB mixers with shorter $\tau_{th}$, the NDR effect on the gain at 1.4 GHz would be more significant.

We should mention here that the gain increase was observed despite the presence of a cryogenic L-band isolator introducing approximately 0.4 dB of loss (factory specification data). The use of the isolator in the NDR regime is indispensable because of the significant impedance mismatch that may affect the LNA noise. It also helps mitigate the parasitic "direct detection" effect due to the background radiation loading, even in HEB devices operating in the PDR.

Further gain improvement requires larger values $\mathcal{L}$, which demands smaller values of $|Z(0)|$. Equation (4) indicates that this requires even smaller $L_B$. For the inductive bias-T, the limit is set by potential leakage of the IF power into the bias circuit. With $L_B$ = 220 nH, the inductive impedance at 1.4 GHz is $X_L$ ≈1.93 kOhm, which is substantially larger than the 50-Ohm input impedance of the LNA. For $L_B$ = 70 nH, $X_L$ ≈ 625 Ohm, which is still more than an order of magnitude greater than 50 Ohm. For smaller $L_B$, $X_L$ would become comparable to 50 Ohm. So, 70 nH is probably nearly the limit to which the bias-T inductance can be pushed for the given IF.

An additional pathway for increasing the gain in the current HEB device is the optimization of the DC-bias resistor $R_B$, which was not addressed in this work. According to Eq. (4), increasing $R_B$ allows for a larger value of $\mathcal{L}$; however, the limiting condition from Eq. (3a) must still be satisfied. Substantial experimental effort will be required to determine the optimal combination of $R_B$ and $L_B$ that maximizes both $\mathcal{L}$ and the conversion gain.

Finally, a more radical approach to mitigating parasitic oscillations is to eliminate the bias inductor $L_B$ altogether. In this inductorless configuration, bias resistor $R_B$ appears in parallel with the HEB and $R_L$, partially shunting the IF signal (Fig. 4). This introduces a multiplication factor of $R_B/(R_L + R_B)$ in Eq. (5a). Besides, the IF impedance presented to the HEB is not $R_L$ only but the parallel connection of $R_L$ and $R_B$, $R_L^* = R_L R_B/(R_L + R_B)$. Hence, the system of Eqs. (5) then becomes:

$$\eta_{no_L} = 2\mathcal{L}^2\left(P_{LO}/P_J\right)\frac{RR_L^*}{\left(R+R_L^*\right)^2}\frac{1}{\left(1+\mathcal{L}\frac{R-R_L^*}{R+R_L^*}\right)^2}\frac{R_B}{R_L+R_B}\frac{1}{1+(\omega_{IF}\tau^*)^2} \quad (6a)$$

$$\tau^* = \tau_{th}/\left(1+\mathcal{L}\frac{R-R_L^*}{R+R_L^*}\right) \quad (6b)$$

We compare the $\eta_{no_L}$ value, assuming $\mathcal{L}$= 2 and $R$ = 5 Ohm with the unshunted $\eta$ value in the PDR regime (Eq. (5a)), where the self-heating parameter is at its highest possible (for example, $\mathcal{L}$ = 0.95) still maintaining the PDR. For $\mathcal{L}$ = 2, Eq. (1) yields $|Z(0)|$ = 15 Ohm; therefore $R_B \approx$ 14 Ohm should be sufficient to provide DC stability. In this case, the ratio of $\eta_{no_L}/\eta \approx$ 3.8 dB. For $\mathcal{L}$= 3, this ratio increases to 5.4 dB. This thought experiment does not consider any variation in the $P_{LO}/P_J$ ratio between the NDR and PDR regimes. Nevertheless, it suggests that the inductorless bias may be a promising approach for achieving a more significant gain enhancement in the NDR regime.

In this work, we have not attempted to obtain system noise temperature data, as this should be preceded by an in-depth study of the output noise in the NDR regime. We also plan to expand the range of $\mathcal{L}$ -parameter values to better understand the full potential of NDR biasing for HEB mixers. Experiments with additional NbN HEB devices are highly desirable, as some may have a narrower superconducting transition and therefore inherently higher $\mathcal{L}$ values. In addition, the receiver optics may need to be upgraded to better mitigate the direct detection effect that is inherent in Y-factor measurements. Similar to a TES direct detector, an IVC with NDR yields higher responsivity, making the parasitic direct-detection effect more problematic.

The potential scale of the noise-temperature reduction can be estimated from the available data for practical THz receivers and from theory [21,22]. Figure 5 shows the double-sideband (DSB) noise temperature of NbN HEBs as a function of radiation frequency $\nu$. Also shown is the absolute theoretical limit arising from the combination of quantum noise, $h\nu/2k_B$, and thermal-energy-fluctuation (TEF) noise ($\approx nT_C/2$) for NbN HEBs [22]. Here, the material-dependent parameter is $n \approx 3.6$ [35] for thin disordered NbN films with $T_C \approx 11$ K.

For example, at $\nu$ = 2.5 THz, the receiver noise temperature is approximately 1,000 K. With a typical optical loss of about 3 dB, the theoretical DSB receiver noise temperature limit is:

$$T_{rec} \approx 2 \times (nT_C/2 + h\nu/2k_B) = 160\text{K}. \quad (7)$$

Thus, the gap between the experimental noise temperature and the theoretical limit (blue dashed line in Fig. 5) is about 840K. Increasing the conversion gain does not affect the TEF noise contribution – because TEF noise scales with the gain – but it should significantly reduce the excess noise arising from Johnson noise and LNA noise. The expected reduction in overall receiver temperature noise temperature (~ 6-times) is substantial, and the practical benefit would be considerable: the mapping speed of an improved mixer could increase by more than an order of magnitude.

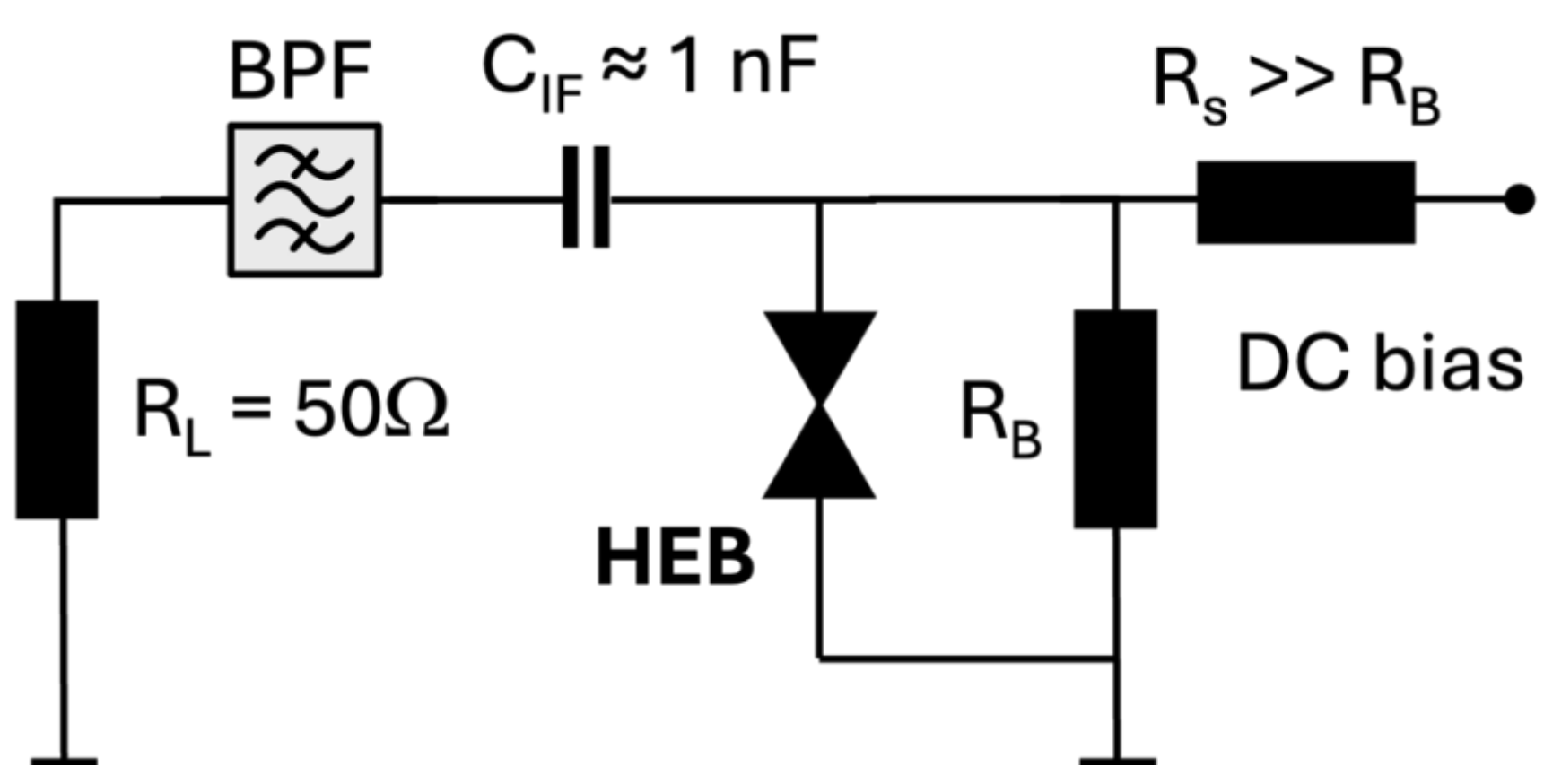

**Figure 4.** Inductorless bias concept for enabling large values of self-heating parameter $\mathcal{L}$. Bias resistor $R_B$ is placed close to the HEB device and its value is smaller than $|Z(0)|$ (see Eq. (1)).

A clearer understanding of the achievable improvement requires broadband measurements of the output IF noise spectrum, which would allow separation of the TEF noise contribution from other noise sources [36].

In conclusion, we have demonstrated an increase in conversion gain in the NbN HEB mixer when it is biased in the negative differential resistance (NDR) regime. This biasing approach should reduce the contributions of Johnson noise and IF-line noise to the overall system noise temperature. Based on the shape of the IVCs, the value of self-heating parameter $\mathcal{L}$ achieved in Figs. 2 and 3 appears to be moderate, slightly exceeding unity. Larger values of $\mathcal{L}$ can lead to larger conversion values. Future work will test the experimental boundary suggested by Eq. (4) and will explore the feasibility of an inductorless DC-bias configuration, in which only the conditions of Eq. (3a) apply, and a significantly larger overall conversion gain may be attainable. The impact of NDR biasing on the intrinsic HEB noise, and therefore on the system noise temperature, represents the next phase of this research and will be addressed in forthcoming studies.

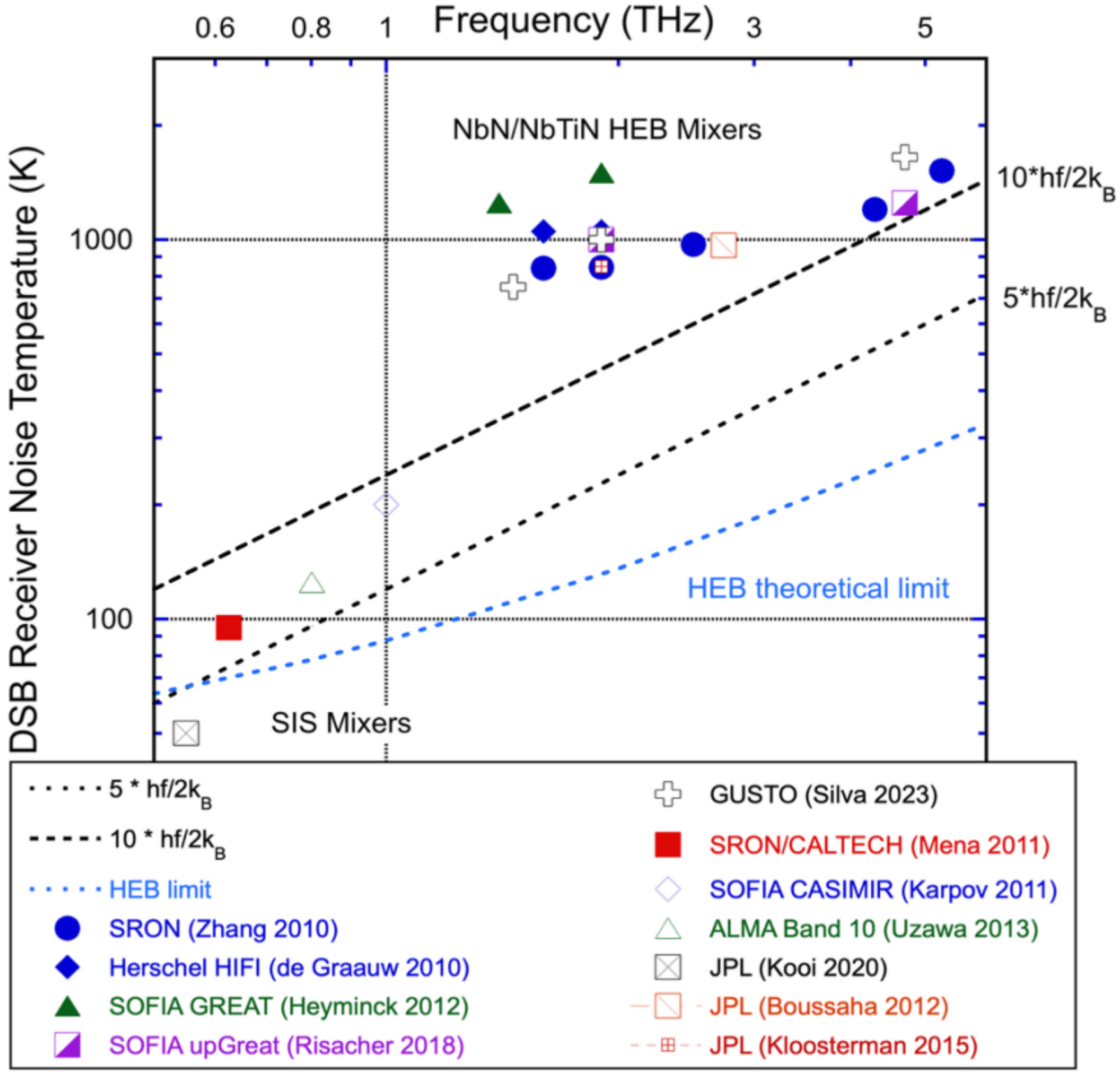

**Figure 5.** Experimental noise temperature in practical HEB receivers and theoretical limits due to the quantum noise and the thermal energy fluctuation (TEF) noise (aka phonon noise).

The research described in this paper was carried out at the Jet Propulsion Laboratory, California Institute of Technology, under a contract with the National Aeronautics and Space Administration. CY acknowledges the support from the NASA Postdoctoral Program.

## AUTHOR DECLARATIONS

### Conflict of Interest

The authors have no conflicts to disclose.

## AUTHOR CONTRIBUTIONS

**Chang Yoo**: Conceptualization (supporting); formal analysis (equal); methodology (equal); investigation (lead); writing/review and editing (equal). **Akim Babenko**: Conceptualization (supporting); formal analysis (equal); methodology (equal), writing/review and editing (equal), funding acquisition (supporting). **Boris Karasik**: Conceptualization (lead); formal analysis (equal); methodology (equal); funding acquisition (lead), project administration (lead), writing/original draft preparation (lead).

## DATA AVAILABILITY

The data that support the findings of this study are available from the corresponding author upon reasonable request.

## REFERENCES

1. M. De Lucia, P. Dal Bo, E. Di Giorgi, T. Lari, C. Puglia, and F. Paolucci, *Instruments* **8**, 47 (2024).
2. B. S. Karasik, A. V. Sergeev, and D. E. Prober, *IEEE Trans. Terahertz Sci. Technol.* **1**, 97 (2011).
3. E.M. Gershenzon, G.N. Gol'tsman, I.G. Gogidze, Y.P. Gousev, A.I. Elant'ev, B.S. Karasik, and A.D. Semenov, *Sverkhprovodimost (KIAE)* **3**, 2143 (1990).
4. W. Zhang, W. Miao, Y. Ren, K.-M. Zhou, and S.-C. Shi, *Superconductivity* **2**, 100009 (2022).
5. K. D. Irwin, *Appl. Phys. Lett.* **66**, 1998 (1995).
6. K.D. Irwin and G.C. Hilton, in *Cryogenic particle detection*, edited by Christian Enss (Springer Berlin Heidelberg, Berlin, Heidelberg, 2005), pp. 63.
7. K.D. Irwin, G.C. Hilton, D.A. Wollman, and J.M. Martinis, *J. Appl. Phys.* **83**, 3978 (1998).
8. H. Ekstrom, B. S. Karasik, E. L. Kollberg, and K. S. Yngvesson, *IEEE Trans. Microw. Theory Techn.* **43**, 938 (1995).
9. R. F. Su, Y. D. Zhang, X. C. Tu, X. Q. Jia, C. H. Zhang, L. Kang, B. B. Jin, W. W. Xu, H. B. Wang, J. Chen, and P. H. Wu, *Supercond. Sci. Technol.* **32**, 105002 (2019).
10. Y. Zhuang and K. S. Yngevesson, *12th Int. Symp. Space THz Technol. (ISSTT 2001)*, 131 (2001).
11. Y. Zhuang and K. S. Yngvesson, *13th Int. Symp. Space THz Technol. (ISSTT 2002)*, 463 (2002).
12. Y. Zhuang, D. Gu, and K. S. Yngvesson, *14th Int. Symp. Space THz Technol. (ISSTT 2003)*, 289 (2003).
13. D. Gu, Y. Zhuang, and S. Yngvesson, *15th Int. Symp. Space THz Technol. (ISSTT 2004)*, 218 (2004).
14. A. Trifonov, C. Y. E. Tong, R. Blundell, S. Ryabchun, and G. Gol'tsman, *IEEE Trans. Appl. Supercond.* **25**, 1 (2015).
15. G.T. Munsterman, *APL Technical Digest* **4**, 2 (1965).
16. R. Morariu, J. Wang, A. C. Cornescu, A. Al-Khalidi, A. Ofiare, J. M. L. Figueiredo, and E. Wasige, *IEEE Trans. Microw. Theory Techn.* **67**, 4332 (2019).

17. C. Kidner, I. Mehdi, J. R. East, and G. I. Haddad, *IEEE Trans. Microw. Theory Techn.* **38**, 864 (1990).

18. L. Wang, J. M. L. Figueiredo, C. N. Ironside, and E. Wasige, *IEEE Trans. Electron Dev.* **58**, 343 (2011).

19. M. E. Hines, *Bell Sys. Tech. J.* **39**, 477 (1960).

20. L. I. Smilen and D. C. Youla, *Proc. IRE* **49**, 1206 (1961).

21. B.S. Karasik and A.I. Elantev, *Proc. 6th Int. Symp. Spc. THz Technol.*, 229 (1995).

22. B. S. Karasik and A. I. Elantiev, *Applied Physics Letters* **68**, 853 (1996).

23. A. I. Elant'ev and B. S. Karasik, *Fiz. Nizk. Temp.* **15**, 675 (1989).

24. B. S. Karasik and W. R. Mcgrath, *Int. J. Infrared & Millim. Waves* **20**, 21 (1999).

25. A. Skalare, W. R. McGrath, B. Bumble, and H. G. LeDuc, *IEEE Trans. Appl. Supercond.* **13**, 160 (2003).

26. C. Yoo, C. Kim, D. P. Cunnane, and B. S. Karasik, (2026), p. arXiv:2601.14463.

27. J. R. Gao, M. Hajenius, Z. Q. Yang, J. J. A. Baselmans, P. Khosropanah, R. Barends, and T. M. Klapwijk, *IEEE Trans. Appl. Supercond.* **17**, 252 (2007).

28. J. R. Gao, J. N. Hovenier, Z. Q. Yang, J. J. A. Baselmans, A. Baryshev, M. Hajenius, T. M. Klapwijk, A. J. L. Adam, T. O. Klaassen, B. S. Williams, S. Kumar, Q. Hu, and J. L. Reno, *Appl. Phys. Lett.* **86** (2005).

29. J. F. Young, B. M. Wood, H. C. Liu, M. Buchanan, D. Landheer, A. J. SpringThorpe, and P. Mandeville, *Appl. Phys. Lett.* **52**, 1398 (1988).

30. H. C. Liu, *Appl. Phys. Lett.* **53**, 485 (1988).

31. D. Wilms Floet, E. Miedema, T. M. Klapwijk, and J. R. Gao, *Appl. Phys. Lett.* **74**, 433 (1999).

32. W. Miao, Y. Delorme, A. Feret, R. Lefevre, B. Lecomte, F. Dauplay, J.-M. Krieg, G. Beaudin, W. Zhang, Y Ren, and S.-C. Shi, *J. Appl. Phys.* **106**, 103909 (2009).

33. A. D. Semenov, R. S. Nebosis, Yu P. Gousev, M. A. Heusinger, and K. F. Renk, *Phys. Rev. B* **52**, 581 (1995).

34. M. Sidorova, A. Semenov, H.-W. Hübers, K. Ilin, M. Siegel, I. Charaev, M. Moshkova, N. Kaurova, G. N. Goltsman, X. Zhang, and A. Schilling, *Phys. Rev. B* **102**, 054501 (2020).

35. Yu. P. Gousev, G. N. Gol'tsman, A. D. Semenov, E. M. Gershenzon, R. S. Nebosis, M. A. Heusinger, and K. F. Renk, *J. Appl. Phys.* **75**, 3695 (1994).

36. H. Ekstrom and B. Karasik, *Appl. Phys. Lett.* **66**, 3212 (1995).